\documentclass[journal]{IEEEtran}
\usepackage{graphicx}
\usepackage{amsmath}
\usepackage{cite}
\usepackage{soul}

\usepackage{atbegshi}
\AtBeginDocument{\AtBeginShipoutNext{\AtBeginShipoutDiscard}}

\usepackage{soul}
\usepackage{color}
\usepackage{lineno}
\usepackage{pifont}

\begin{document}

\title{Computational Compressed Sensing of Fiber Bragg Gratings}

\author{Srikanth Sugavanam*, Adenowo Gbadebo, Morteza Kamalian-Kopae, Angshul Majumdar}

\thanks{*Srikanth Sugavanam is with the School of Computing and Electrical Engineering, IIT Mandi, Kamand, Himachal Pradesh - 175075. Email - ssrikanth@iitmandi.ac.in}
\thanks{Adenowo Gbadebo and Morteza Kamalian are  with the Aston Institute of Photonic Technologies, Aston University, Aston Triangle, Birmingham B4 7ET, United Kingdom.}
\thanks{Angshul Majumdar is with the Indraprastha Institute of Information Technology, Okhla Industrial Estate, Phase III, New Delhi}

\maketitle

\begin{abstract}
State-of-the-art fiber Bragg grating interrogators utilize mature concepts and technologies like tunable lasers, optical spectrum analyzers and a combination of time, wavelength, or spatial division demultiplexing approaches. Here, we propose the use of  computational compressed sensing (CS) techniques for interrogation of fiber Bragg gratings, reducing interrogator complexity by using a broadband ASE light source and single-pixel detection. We demonstrate temperature sensing using a pre-calibration approach to achieve reconstruction accuracy comparable to uncompressed measurements. We extend these principles for the interrogation of sensor networks, presenting strategies for tackling sparsity considerations. Our proof-of-principle demonstrations show how the presented computational compressed sensing techniques can provide an alternative for realizing low-complexity, small footprint interrogator configurations.

\end{abstract}


\section{Introduction}

State of the art fiber optic sensors based on fiber Bragg gratings (FBGs) are commercially deployed solutions that are routinely used in various environments and applications \cite{CHAN2006648,chiavaioli2017biosensing,Caucheteur2015,DiSante2015Aircraft}. The basic sensing element in these configurations is the fiber Bragg grating, which can be inscribed in a host of conventional and specialized fibers \cite{Gillooly2011}.  The intrinsic material response to external stimuli like temperature or stress (or those which are transduced to these parameters as such) can manifest as a shift of the FBG central wavelength, an effect that has been used ingeniously in the realization of a host of sensors (see for e.g. \cite{othonos2006fibre,cusano2011fiber}). 

As conventional FBGs are typically a few millimeters or centimeters long, the observations are highly localized. This can be a detriment when a more distributed sensing approach is called for \cite{thevenaz2006review,DistributedReviewMasoudi2011,soto2016intensifying,hartog2017introduction}. Indeed, non-FBG-based distributed sensing configurations invoking 
linear principles like OTDR and OFDR, or 
nonlinear gain mechanisms like SBS or SRS have demonstrated high sensitivities, with dependable commercial viability and deployed as such. There the enhancement of spatial resolution and sensitivity come at the cost of increased configuration complexity. 
Alternatively, several FBGs can be spliced together or inscribed into long spans of fibers, sometimes running into several tens of kilometers to realize quasi-distributed sensing configurations or networks \cite{fernandez2012optical}. These offer complementary solutions to distributed fiber sensors, striking a trade-off between performance and complexity. 

The shifts of the central wavelengths of FBGs and their networks are obtained using fiber Bragg interrogators. This is a collective term for a device or a combination of devices that realize this function of extracting the spectral shift of the FBG sensors. Several modalities of interrogators exist, which are based on principles of time \cite{Chan}, wavelength \cite{Jackson:93,Norman2005}, or space-division multiplexing, and/or their combinations \cite{Rao:95,Rao:96}. The development of robust swept sources in turn have led not only to simpler time-domain implementations of FBG-based sensor networks, but also enhanced the functionality of distributed sensing solutions.

Indeed, the interrogators based on these technologies are extremely reliable, and the real-world use of such technologies are well-established. But their implementations can get quite complex. For instance, telecom-grade arrayed waveguides combined with simple peak-detection algorithms allow for simpler demultiplexing architectures, but need either FPGA-based switching or multiple detectors for interrogation of the individual channels \cite{Hegyi}. Demultiplexing schemes based on OTDR coupled with signal processing techniques have also been demonstrated. For instance, weak FBGs (1-5\% reflectivities) with nominally similar central wavelengths inscribed over several tens of kilometers of fiber have shown the potential for ultralong sensing \cite{Wang:12,Wang}. In their case, the complexity lies in the inscription of the gratings, and in their interrogation that is facilitated by wavelength tuning of the source.

In this context, an area of research that is gaining rapid prominence is that of single pixel detection techniques. It exploits the fact that many real world signals can be projected on a basis that can yield a more compact representation, e.g. using only a few leading terms. These techniques use a combination of experimental and computational techniques for acquisition of complex, multimodal signals and their reconstruction. They yield simplified measurement configurations, often alleviating the need of sensitive complex optics, and replacing expensive detector arrays with a single-pixel detector. While initially demonstrated for imaging applications \cite{DuarteSinglePixel, edgar2019principles}, the approaches are now being increasingly translated and explored for the realization of compact interrogation platforms for diverse applications like spectroscopy \cite{Wang:14, starling2016compressive, Soldevila:19}, generation and compression of sparse radio frequency signals \cite{bosworth2013high,Zhu:18}, high speed imaging \cite{bosworth2015high,han2016imaging}, and more recently, even for hyperspectral imaging \cite{wang2019single}. Indeed, the application of compressed sensing principles has been investigated for FBG interrogation, specifically using incoherent OFDR techniques \cite{WerzConf,WerzMDPI}. However, the interrogation methodology there was quite complex, requiring the use of tunable lasers and vector network analyzers.

Here, we draw on the above-stated principles of computational single-pixel imaging, and propose the use of a single-pixel based, computational compressed sensing approach for interrogation of a fiber Bragg grating-based sensor network \cite{sugavanam2019compressed,SriSPIE2020}. This capitalizes on the underlying sparsity of the measurand, i.e. the discretely sampled FBG reflectance spectra. The use of an ASE-based optical source and single-pixel detection reduces the complexity of the interrogator optical configuration. Our measurements are performed directly in the optical spectral domain, which lead to the simpler experimental configuration. The spectra are reconstructed using computationally light algorithms, which can potentially be implemented on compact microcomputer platforms. Here, we use a programmable filter for demonstration of proof-of-principle, but in essence this can be replaced using a bank of pre-fabricated filters. 
Thus the resulting interrogator configuration can be constructed to have a small footprint comprising of almost no mechanical moving parts like etalons. The proof-of-principle configuration demonstrated by us shows the viabilty of using CS-based approaches as a low-cost, low-complexity alternative for state-of-art FBG interrogators.

In the following, we first demonstrate the principle of compressed sensing as applied to FBG interrogation. We derive expressions that relate the efficacy of reconstruction to measurement precision. We then demonstrate its use in a temperature sensing configuration, where we use a pre-calibration approach to achieve reconstruction in the compressed regime. Finally, we extend these principles for demonstrating the interrogation of a multiplexed FBG sensor network, highlighting the challenges therein in the process.

\section{Results}
\label{sec:Results}

\subsection{Compressed sensing (CS) principle}

Compressed sensing (CS) is a convex optimization-based reconstruction methodology that exploits the intrinsic sparsity of the signal. The mathematical formulation of CS states that if the signal is sparse in some basis, it can be reconstructed by projecting it on a basis that is completely incoherent to the sparse basis using $l_1$ minimization, where a unique solution can be guaranteed. 

Let $\mathbf{y} = \{y_1, y_2, ...y_n\}$ denote the signal of interest of dimension $n$, which has a sparse representation of the form $\mathbf{y'}$ of the same dimensionality $n$ in a different basis. Let $\mathbf{A}_{m\times n}$ denote the sampling operation performed $m$ times on the signal $y$, in a way that the $m$ sampling configurations are completely uncorrelated to each other. The measurand $x_m$ of interest in CS is then the integrated signal arriving from the sampler. That is, 
\begin{equation}
\sum_{n}A_{mn}y_n = x_m.
\end{equation}\label{Eq:ExpMeas}

If the conditions of CS are met, then the coefficients $y’$ can be estimated by solving the convex optimization problem
\begin{equation}
   arg min{\left\lVert \mathbf{y'}\right\rVert}_1, s.t. \mathbf{A}.\mathbf{y} = \mathbf{x}.
\end{equation}\label{Eq:CS}

If the conditions of sparsity and incoherent sampling highlighted above as satisfied, the underlying mathematical framework guarantees that the solution obtained is unique \cite{candesIEEE, candes2008restricted}(also see \cite{edgar2019principles}). Further, the theory states that such a reconstruction can be realized by making fewer measurements than the dimension of the signal itself, i.e., the sensing is compressed ($m<n$). A compression factor can then be defined by the ratio $n/m$.  This will be achieved as long as the basic requirements of signal sparsity and sampling incoherence are satisfied \cite{candes2007sparsity}. Further no a priori information about the signal is required, i.e. the signal sampling can be non-adaptive. 

We now show how the CS-approach can be adapted for enabling FBG interrogation.

\subsection{Experimental configuration}

\begin{figure}[t]
\centering
\includegraphics[width=\linewidth]{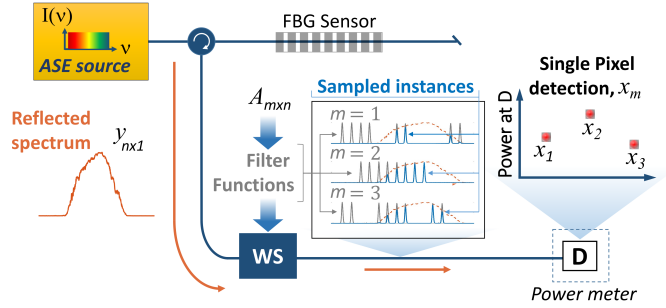}
\caption{\textbf{Compressed sensing (CS)-enabled Fiber Bragg Grating interrogation.} Schematic representation of experimental configuration. The reflected spectrum $\mathbf{y}_{n\times 1}$ is sampled $m$ times by the measurement matrix $\mathbf{A}_{m\times n}$. The shown sampled instances are experimentally obtained. These instances are spectrally integrated by the single-pixel detector \textbf{D} to yield the power values $x_m$.  ASE - amplified spontaneous emission, FBG - Fiber Bragg Grating, WS - Programmable filter.}
\label{fig:ExpConfig}
\end{figure}

\begin{figure*}[t]
\centering
\includegraphics[width=\linewidth]{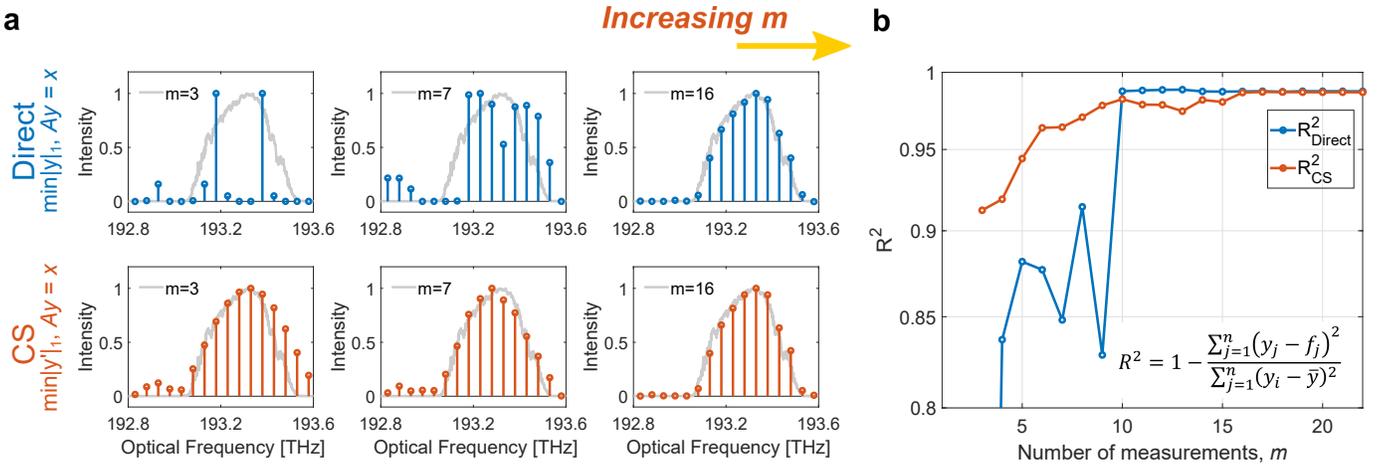}
\caption{\textbf{Compressed sensing (CS)-enabled Fiber Bragg Grating interrogation.} \textbf{a.} Reconstruction from single pixel measurements using the direct, equation solving approach (top row), and CS-based approach (bottom row). Gray curves - grating spectra as measured independently using a high resolution OSA. CS reconstructions can be obtained with a limited number of measurements $m$ smaller than the actual number of points $n$ being reconstructed. \textbf{b.} Reconstruction efficacy for the direct and CS approaches quantified in terms of the coefficient of determination, $R^2$. For $m \geq 10$, $R^2>0.98$ for both the CS and direct approaches.}
\label{fig:Results}
\end{figure*}

Figure \ref{fig:ExpConfig} shows the experimental configuration used for CS-based FBG interrogation.
The fiber Bragg grating has a Gaussian spectral profile of width  $\sim$ 400 GHz (2 nm). The light source is an erbium-doped fiber amplifier generating broadband ASE (total output power$<$50 mW). In place of the conventionally used spectrometer, we use a programmable optical filter (Finisar 4000s) and photodetector/power-meter combination for FBG interrogation (Anritsu ML910A). In addition, we used a high resolution optical spectrum analyzer (Finisar WaveAnalyzer 1500S, resolution bandwidth 1.4 pm near 1551 nm) for verification of experimental results. These components enable the CS-based approach for random sampling of the reflected spectrum and its integrated detection. As per requirement of the CS-methodology, the random sampling functions must be uncorrelated. To this end, we use the Hadamard basis to realize the measurement matrix $\mathbf{A}_{m\times n}$, whose rows are known to be uncorrelated
(see for example, \cite{satat2017lensless}). Using the programmable filter, we design frequency combs 
where the heights of the individual peaks are then made to take the values ${A}_{mn}$, effectively turning them on or off. This results in filter functions as shown in the inset of Fig. \ref{fig:ExpConfig}, which randomly sample the grating profile. Here, the sampling grid is chosen to have a discretization of 50 GHz.
The low-bandwidth power meter then functions as our single pixel detector, and is used to measure the spectrally integrated intensity $x_m$ for each randomly sampled instance. These randomly sampled instances can now be used to reconstruct the grating reflectance spectrum by using an appropriate minimization routine. We used a standalone computer (Dell Inspiron 7537) to automate the data acquisition and spectral reconstruction.

\subsection{CS-based spectral reconstruction}

The $l_1$ norm minimization is implemented numerically using the open-source CVX package \cite{cvx}. The inputs to the optimization routine are the single-pixel power values $\textbf{x}$ and the measurement matrix $\textbf{A}$. Here, we use a computational compressed sensing approach for the reconstruction: instead of using the actual filter transmittance functions, we use the computationally generated Hadamard matrices for the reconstruction. This gives us a computational advantage - while the actual filter transmittances have a dimensionality of $n\sim50,000$, the computational matrices have $n = 16$. The optimization routine then reconstructs the reflectance spectra only for the sampled locations, and not for the zeros in between, which leads to faster convergence times. To the best of our knowledge, this is the first time such a computational compressed sensing approach -  specifically where the sampling is performed directly in the spectral domain, and the computational matrices are used for the reconstruction rather than the actual experimentally measured filter profiles - has been exploited for FBG interrogation.

The object of our interest are the spectral intensities as sampled by the individual peaks of the programmable filter. However, here the sampling is already made in the spectral domain. To effect CS, we need to move to a different basis. Here, we choose to move to the Fourier domain, and use the following reconstruction strategy:

\begin{equation}
arg min{\left\lVert \mathbf{y'}\right\rVert}_1, s.t. \mathbf{A.y} = \mathbf{x}, 
\end{equation} \label{Eq:CS_FFT}
where 
\begin{equation}
\mathbf{y'} \equiv FFT^{-1}(\mathbf{y}).     
\end{equation}\label{Eq:CS_FFT2}

That is, we numerically move to the Fourier domain, and search for a solution whose Fourier coefficients have the smallest $l_1$ norm. This is verified automatically by the optimization routine when we define the objective function to be minimization of the $l_1$ norm of the FFT of $\textbf{y}$. Presently, we have used the default tolerance level of CVX to determine convergence. Once we arrive this solution, we can then use an inverse FFT operation to recover the actual spectral samples. We note that a similar combination of an ASE source and programmable filter was used recently for sparse recovery of time-domain narrowband radio signals \cite{Zhu:18}. However there the sampling was performed in the complementary optical spectral domain, where the signal representation was sparse. In our work, the sampling is performed in the measurement domain of interest itself, while the projection to the sparse Fourier domain is achieved numerically. Also, our approach is very similar to the spectral sampling methodology used by Starling et al \cite{starling2016compressive} for the reconstruction of discharge lamp spectra. There, the optimization condition was based on the minimization of the total variation. Here, our use of $l_1$-norm minimization is seen to be viable, as we have numerically moved to the Fourier domain. In addition, our approach is also different from the model-based compressed sensing approach \cite{WerzConf} and other more advanced model-based approaches demonstrated for FBG interrogation. There, the FBGs were interrogated with incoherent OFDR that employed a tunable laser and a vector network analyzer for measuring the complex valued radio-frequency response parameter $s_{21}$. The grating reflectivities were modeled as complex valued parameters, which were then extracted using a least-squares algorithm. In contrast, in our work we directly interrogate our FBG in the spectral domain. Our interrogation is also simplified due to the use of the ASE source, a single pixel photodetector, and a programmable filter which can be replaced by a filter bank. Further, the $l_1$-norm minimization condition directly yields the grating reflectance without the need for use of any models, or to that end, any a priori knowledge of the grating profiles. Our proposed technology thus offers a low-complexity complementary modality for CS-based interrogation of FBG gratings.

To check the effectiveness of the Fourier approach, we compare with the reconstruction obtained directly in the spectral/sampling domain, i.e. as per the methodology 
\begin{equation}
arg min{\left\lVert \mathbf{y}\right\rVert}_1, s.t. \textbf{A}.\textbf{y} = \textbf{x}.
\end{equation}\label{Eq:Direct}
We call this the direct approach, as Eq. 5 amounts to solution of an ill-conditioned system of equations when we are in the compressed regime, i.e. $m<n$. The primary difference between this and the CS approach of Eq. 3 is now the FFT operation on $\textbf{y}$ indicated above is not carried out. Instead, the objective function now is the minimization of the $l_1$-norm of y itself.

Figure \ref{fig:Results}a shows the reconstructions of the reflectance spectra as obtained with the direct domain (top row) and CS approaches (bottom row). The stem plots are indicative of the smaller matrix used by us for the reconstruction. Immediately, we see that the CS-based reconstructions are much better than the direct domain approach, even for a very small number of measurements $m$. The reconstructions improve progressively as $m$ increases. In both approaches, we order the measurements used for the reconstruction in the order of decreasing value of $x_m$, as this is seen to yield the fastest convergence. These results show that the computational approach indeed is a viable strategy for the reconstruction of the reflectance spectra. 

The efficacy of the reconstructions can be quantified using the coefficient of determination $R^2$ \cite{devore2011probability}, defined as
\begin{equation}
    R^2 = 1-\frac{\sum_{j = 1}^{n}\left(y_j - f_j\right)^2}{\sum_{j = 1}^{n}\left(y_j - \bar{y}\right)^2},
\end{equation}
where $y_i$ are reconstructed spectral samples, $\bar{y}$ is their mean, and $f_j$ are the corresponding values of the reflected spectra directly measured using the high resolution OSA. $R^2=1$ indicates perfect reconstruction. 

Fig. \ref{fig:Results}b give the $R^2$ values for direct domain and CS-apporaches, and it is evident that $R^2$ is higher for the CS-based reconstruction than the direct sample domain reconstruction, improving almost monotonically with $m$.
The sudden jump at $m=10$ for the direct domain reconstruction is due to the transformation of the optimization problem Eq. 5 to the solution of a well-conditioned system of equations, when the number of equations equals the number of unknowns, i.e. the number of peaks under the FBG profile. 
This effect can also be seen for the CS approach, where the value of $R^2$ becomes constant for $m>=16$, and approaches the results obtained for the direct modality. This is again because now $m$ equals the size of the frequency grid used for the sampling, and the system becomes determinate. The results show that the CS-approach gives a better reconstruction in the compressed regime $m<n$, showing the viability of the CS-approach for fiber Bragg grating interrogation. 

\subsection{Centroid detection and measurement precision}

\begin{figure}
\centering
\includegraphics[width=\linewidth]{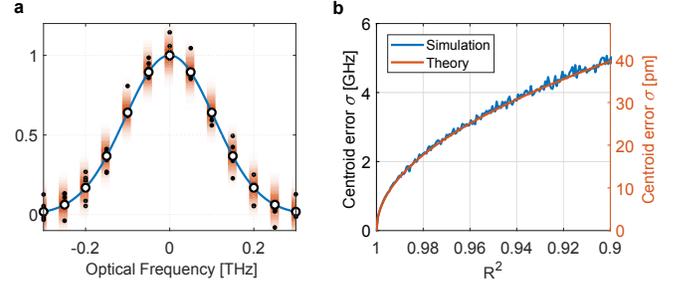}
\caption{\textbf{Numerical estimation of centroid precision as a function of $R^2$} \textbf{a.} Gaussian reflection profile used for the simulation. Reconstruction errors are emulated as Gaussian noise (orange shaded region about sampled points). Some of the randomly generated samples are shown as black dots, shown here for $\sigma_y = 0.01$. \textbf{b.} Centroid precision as a function of $R^2$, with good agreement between theory and simulation (ensemble size - $10^3$ spectra).}
\label{fig:Simulation}
\end{figure}

The reconstructions of Figure \ref{fig:Results}a can be used to measure the spectral locations and shifts of the FBG using techniques like centroid detection, Gaussian or polynomial curve fitting \cite{tosi2017review}. 
Here, we use centroid detection as it allows for a straightforward comparison of theoretical estimations with our experimental results. Centroid detection 
gives the centre of mass of the measured reflectance profile optical frequency $\nu_c$ from the reconstructed spectral intensities $I_j$ at the optical frequencies $\nu_j$ as
\begin{equation}
    \nu_c = \sum_{j=1}^{n} \nu_j.y_j/\sum_{j=1}^{n} y_j.
\end{equation}\label{eq:centroid}
For a grating with a Gaussian profile, the centroid $\nu_c$ is simply the location of the central frequency of the grating profile. By striking a trade-off with accuracy, it is possible to determine the centroid using only a limited number of sample points $n$. 

The centroid precision will also depend $R^2$.
Using the definition of the centroid above and basic error propagation theory \cite{tellinghuisen2001statistical}, we can estimate the centroid variance, $\sigma^2_{\nu_c}$, as a function of $R^2$:
\begin{equation}
\sigma^2_{\nu_c} = \sigma^2_y\frac{\sum_{j=1}^n\left(\nu_j-\nu_c\right)^2}{\left(\sum_{j=1}^n y_j\right)^2},
\end{equation}\label{eq:centroiderror}
where 
\begin{equation}
    \sigma_y^2 = \frac{1}{n}\sum_{j=1}^{n}\left(y_j - f_j\right)^2 
\end{equation}
and
\begin{equation}
    R^2 = 1 - \frac{n\sigma_y^2}{\sum_{j=1}^{n}\left(y_j - \bar{y}\right)^2}.
\end{equation}

The expressions can be used to arrive at a nomograph like Fig. \ref{fig:Simulation}b, which can help ascertain the value of $R^2$, and hence the number of samples $m$ required to attain the required level of centroid precision. The blue curves show the corresponding result obtained via a  simulation of noise superimposed on a grating reflectance profile, employing $10^3$ spectra for different values of $R^2$. In the example shown, the sampling conditions are similar to the experiment (Fig. 2).  
The agreement between theoretical and simulated results confirm our understanding of the relation between reconstruction efficacy and centroid variance, assuring the correctness of using such an approach to determine interrogator resolution. 

Fig. \ref{fig:Simulation} indicates that in our case, for $m < n$, we can achieve a centroid variance $\sim$20 pm, reaching  to $\sim$ 10 pm for the uncompressed regime. We specifically chose centroid detection, as while it offers a poorer resolution, here it allows for a simple and direct verification of our analytical formulae. Indeed, conventional OSAs based on scanning filters (see for e.g. \cite{Ibsens}), and allied regression-based curve-fitting algorithms offer lower uncertainties \cite{tosi2017review}. 
Nevertheless, we have shown above that our single-pixel based CS approach can offer a comparable performance, where the values can be improved by sampling more points under the FBG, and thus is currently limited by the programmable filter we use. Our demonstrated approach thus presents an alternative to conventional interrogation modalities, should a lower complexity configuration with a smaller footprint be desired.

\subsection{CS-based Temperature sensing}

\begin{figure}[t]
\centering
\includegraphics[width=\linewidth]{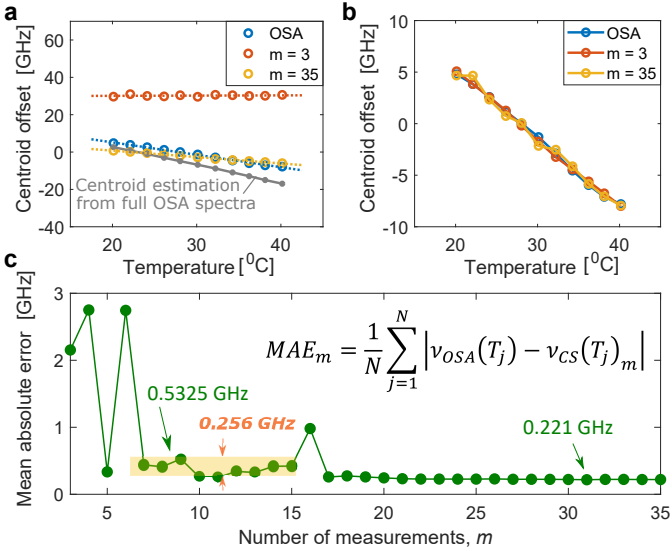}
\caption{\textbf{CS-enabled Temperature sensing using FBGs.} \textbf{a.} Grating centroid as a function of temperature obtained using $n=16$ samples from high-resolution OSA (blue) and CS approaches for $m=5$ (yellow) and $m=35$ (orange) measurements. Dotted lines are corresponding linear fits. Gray plot shows centroid values estimated from full ($n\sim50,000$) OSA spectra. 
\textbf{b.} Grating centroid vs. temperature after calibration. 
\textbf{c.} Mean absolute error, $MAE_m$, of the centroid values in Fig. \textbf{b} averaged over measured temperatures, as a function of $m$. The orange shaded regions correspond to $m<n$, showing pre-calibration brings down the MAE, enabling compressed sensing.}
\label{fig:CSTempSense}
\end{figure}

We now apply the CS-based interrogation in a temperature sensing configuration. The experimental configuration is the same as shown earlier, where the FBG is now mounted on a Peltier element. 

As analytically estimated by Eq. (8), the centroid precision falls in the range between 10 and 20 pm for our configuration, the FBG is heated to give rise to centroid shifts greater than this uncertainty. Fig. \ref{fig:CSTempSense}a shows a comparison between the centroid values derived for different temperatures using the CS-approach for two values of $m$, and that obtained at the same time of measurement using an independent high resolution OSA. Specifically, we use a sampled version of the direct OSA spectrum to enable proper comparison, comprising of n = 16 points corresponding to the reconstructed spectral locations. Dotted lines are their corresponding linear fits. Here, the centroid shifts are represented as offsets from an arbitrarily chosen reference optical frequency (treated as a constant for all the measurements). Here we see that there is an error between the actual and reconstructed values, which are systematic. 

For a given value of $m$, the reconstruction preserves the monotonic relationship between temperature and centroid shift. This behavior is observed for all values of $m$. It is thus possible to systematically correct for the reconstruction errors in the compressed regime via a pre-calibration procedure. Specifically, the linear fit parameters of the temperature calibration curves Fig. \ref{fig:CSTempSense}a for different values of $m$ can be obtained before-hand, and can be compared to the fit obtained for uncompressed or independent direct OSA measurements to obtain the slope and intercept correction factors. Fig. \ref{fig:CSTempSense}b shows the result of this correction, which shows a very good agreement between actual and reconstructed values. The temperature sensitivity is seen to be around 5 pm/$^0$C. Theoretical sensitivities for conventional FBGs are closer to 10 pm/$^0$C \cite{zhang2019multichannel}. 
The lower sensitivity in our case can be attributed to the interplay of two factors -  estimation of centroids from a sampling of an asymmetric grating, and the systematic change of grating parameters with temperature. For an asymmetric grating spectrum, the estimate of the centroid value depends on the extent of sampling. This is shown in Fig. 4a, where the bold gray plot shows the centroid value estimated from the direct OSA spectra ($n\sim50,000$). However, the estimates of the centroid values change when the sampled spectra are used ($n=16$), resulting in the observed lower sensitivity. In addition to this effect, we have observed a systematic change of the grating parameters, viz. the FWHM and its skewness, which while nominal, further contribute to the observed lowering of the sensitivity.

The lower sensitivity here does not compromise the presented results and our primary purpose. That is, we have shown that the CS-approach, together with pre-calibration, can be used to obtain compressed spectral measurements. Specifically, we have shown that our methods do indeed resolve centroid shifts to the precision as analytically estimated using Eq. 8. The sensor sensitivity can be improved further by using a finer sampling comb (i.e. larger value of $n$), using better fitting procedures, or by using sensitized grating configurations. The temperature resolution is ultimately limited by the reconstruction efficacy afforded by the CS-methodology used.

The uncertainty in the measured centroid value can be further quantified using the mean absolute error, as shown in Fig. \ref{fig:CSTempSense}c. Here, $\nu_{OSA}(T_j)$ correspond to the values of the centroid as estimated from a linear fit of the direct OSA measurements, and $\nu_{CS}(T_j)_m$ are the values of the centroid as obtained using the CS methodology at the temperature $T_j (j = 1,2,...,N)$ using $m$ sampling measurements. It shows that the error for small values of $m<n$ (i.e. CS regime) is very close to those obtained for $m\geq n$ (i.e. uncompressed measurement). Thus, we have shown that true compressed sensing is possible using a pre-calibration approach ($m<n$). Here the compression value is 16/5 $\approx$ 3, or 16/7 $\approx$ 2.
Indeed, we note smaller values of $m$ have lower precision even after calibration, yet this is significantly lower than those before pre-calibration (see Fig. \ref{fig:Results}b for example). To the best of our knowledge, the systematic bias and offset noted in our configuration and its use for error compensation has not been reported elsewhere, and it would be interesting to investigate this effect in other similar CS-based configurations. 

The dynamic measurement capabilities of the configuration as demonstrated here is restrictive.  While state-of-the-art FBG interrogators based for e.g. on scanning filters or sources (e.g. \cite{Ibsens}) can provide acquisition speeds of several kilohertz, the acquisition speed in our case is primarily limited by the switching speed of the programmable filter. In principle, this speed bottleneck can be overcome by using a bank of pre-fabricated filters,where the compressed measurements ($m<n$) can be made simultaneously using multiple detectors. In addition, computationally faster algorithms can be used to reduce the processing speeds further. Optimization of the interrogator design towards dynamic acquisition will require a bottom-up engineering approach, and is thus reserved to future work.

\subsection{CS-based Demultiplexing}

\begin{figure*}[h]
\centering
\includegraphics[width=0.9\linewidth]{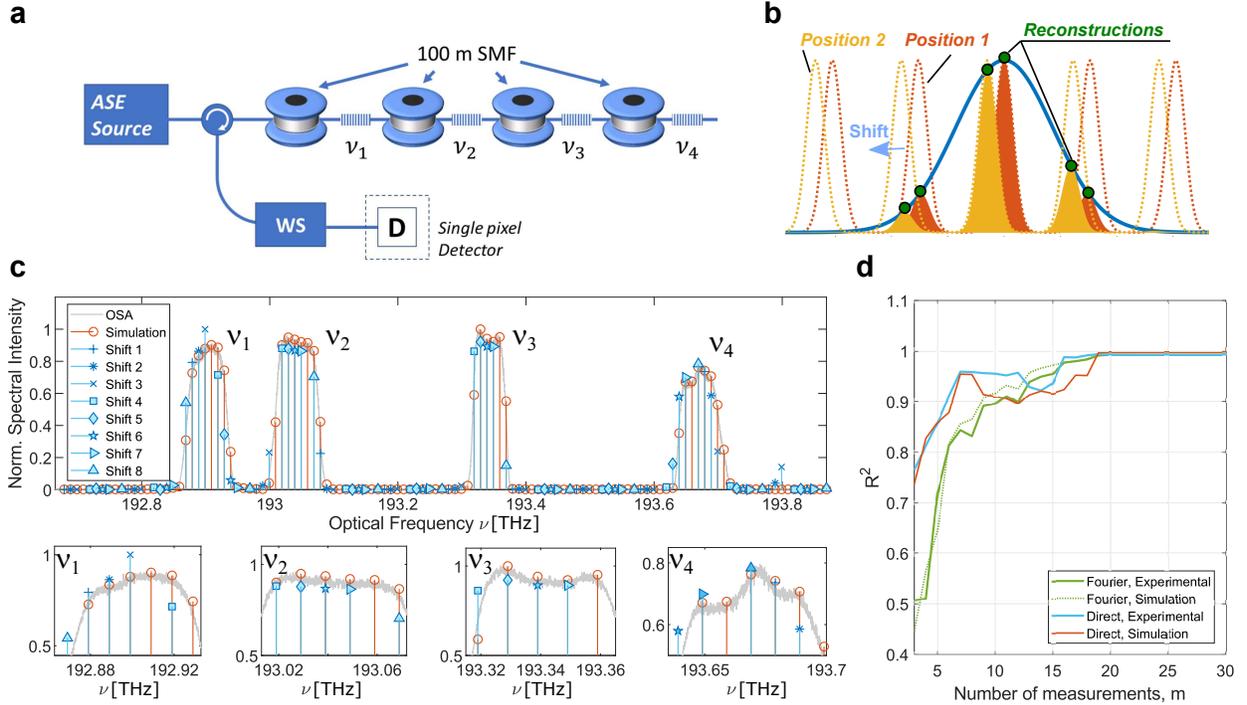}
\caption{\textbf{CS-enabled demultiplexing of sensor networks} \textbf{a.} Experimental configuration, where $\nu_1, \nu_2...$ indicate central frequencies of the gratings. \textbf{b.} Principle of sampling. The sampling grid is shifted sequentially when the random measurements are made. \textbf{c.} Single-pixel detection based spectral characterization of the multiplexed FBG network. Reconstruction was performed using the direct domain approach (Eq. 5). Each blue-shaded stem plot ($+, *,$ \ding{72}, \ding{115}, ...) represents a different shifted location of the sampling grid. Plot indicated with orange open circles are results of simulation that employed the directly measured network profile (gray curve). Sub-figures show close-ups of the reconstructed profiles. \textbf{d.} Efficacy of reconstruction $R^2$ for the full reconstruction. It also shows the comparison between the results obtained using the CS (Fourier) approach, indicating here a better performance for the direct domain approach.}
\label{fig:CSMultiplex}
\end{figure*}

Insofar, we have shown that the CS-approach can be viably used for interrogating individual gratings. We now consider strategies for interrogating sensor networks. Figure \ref{fig:CSMultiplex}a shows an experimental configuration where four FBGs with different central frequencies are spliced 100 meters apart. Each FBG has a nominal spectral width of 0.2-0.4 nm, which is narrower than the FBG used in the previous sections, and is closer to those used typically in structural sensing and other real-world applications. Such sensor networks are typically interrogated using wavelength-division multiplexing techniques.

The challenge in using CS-approaches for such configurations is two-fold. Firstly, there is a limit to the number of points that can be sampled under the reflectance spectrum of each grating owing to available resolution on the programmable filter. Further, the addition of multiple gratings affects the sparsity of the reflected signal, which will affect the reconstruction efficacy. To tackle both these challenges simultaneously, here we adopt a strategy wherein the reflected spectrum is sampled in a very sparse fashion, to the extent of only one or two points per grating. Several random measurements based on this sampling grid are then used to reconstruct this particular instance of sparse sampling of the reflectance spectrum. Next, the sampling grid is itself shifted to yield the reconstruction at a different spectral locations across the reflected spectrum. This step-and-shift process is then repeated several times to acquire the requisite samples to reconstruct the reflectance spectrum (see Fig. \ref{fig:CSMultiplex}b). At present, we sample this using one programmable filter, and process the results using an off-board computer. Hence this process takes a relatively long-time ($\sim$few minutes). This can be reduced in principle by using dedicated hardware and on-board processing.  

Fig. \ref{fig:CSMultiplex}c shows the experimental results of this reconstruction (blue stem plots) obtained using a filter grid of discretization 100 GHz, where the grid is stepped 8 times by 10 GHz. This five-fold betterment in the discretization from the previous sections (50 GHz) results in a cumulative sampling grid that measures more points under the reflection spectrum. For reconstruction, like previously we proceed to order the measurements in order of decreasing optical power values. Towards this, we use a priori information we have about the grating configuration by choosing the ordered measurement matrix corresponding to the shifted grid location that samples the most gratings in the network, and use this matrix and ordering of $x_m$ for reconstruction of all the shifted locations. In this approach, we note that we obtain reconstruction convergence for all the locations. In fact, we have examined that there are nominal differences in the convergence characteristics from those obtained using the individual ordered matrices. For the presented configuration, this greatly simplifies the reconstruction approach, requiring the use of only one ordered matrix for the interrogation of all gratings in the multiplexed network. 

The reconstructions shown in Fig. \ref{fig:CSMultiplex}c are obtained using 19 measurements for each reconstruction. The reconstructions are obtained using the direct approach (Eq. 5). The orange stem plots show for comparison the reconstructions obtained with a numerical simulation under conditions same as that of the experiment, where the sampling filter shapes were simulated by convolving the ideal filter response with the measurement matrix $\textbf{A}_{mn}$, and multiplying these with the directly measured high resolution spectra (grey plot), as per Eq. 1. This simulation shows the feasibility of using direct domain reconstruction for interrogating multiplexed networks, whilst also revealing residual errors for the experimental reconstruction. At present, we believe we are limited by the programmable filter we used; for instance, as we operate the instrument at its specified limits (i.e. narrowest filter comb possible, smallest shift possible, etc.) we noted we could not obtain faithful reproduction of the sampling matrix for one of the shift locations. It is expected that this can be improved by the use of finer sampling modalities when they become available.

Fig. \ref{fig:CSMultiplex}d shows the efficacy $R^2$ for the full reconstruction obtained using both CS and direct domain approaches (numerics and experiment), resulting in a reconstruction efficacy of $R^2> 0.98$  for the full reconstruction. The results show that the direct sample domain reconstruction performs slightly better, especially when the sparsity is reduced. This can be attributed to increased complexity of the signal being sampled, which does not result in a compact representation in the basis domain. 

The cumulative sampling grid here corresponds to $n = 120$. Here, we have obtained a reconstruction using $m = 19 \times 8 = 152$ measurements, moving us away from the compressed regime. Indeed, this limitation is could be primarily due to the increase in complexity owing to the use of additional gratings. This calls for measurement trade-offs, for instance using a coarser sampling grid at the cost of making more shifts. In contrast, the incoherent OFDR-model based CS method \cite{WerzConf} has been used successfully for demultiplexing of ten gratings, and even up to 20 gratings using the advanced model-based approaches \cite{WerzMDPI} .  In our work, in spite of the current limitations, the measurement configuration still remains less complex than other complementary methodologies, which could be attractive for low-cost applications. Another reason could be our use of the simple optimization methodology as given in Eq. 5. Alternative basis spaces, and similarly more case-specific optimization/regularization conditions can be explored for addressing these limitations towards addressing the challenges encountered above, and for achieving compressed sensing in the multiplexed configuration.

\section{Conclusion}
Optical sensors occupy a significant share of the photonics market, bringing in millions of dollars in annual revenue. It is thus evident that improvement of the performance of the interrogators used, along with reduced procurement and maintenance costs can result in significant dividends. The demonstration of viability of a computational compressed sensing approach for FBG sensor interrogation is a step in this direction. 

In the presented work here, we have shown how we can realize a low complexity, computational compressed sensing approach to recover the grating reflectance profile from random samplings of it performed directly in the spectral domain, and a single point detector. This single-pixel approach alleviates the need of spectrometer-based configurations. 
The most complex component of the demonstrated configuration is the programmable filter. However, as we have demonstrated, relatively few measurements are needed for the reconstruction, and hence the programmable filter can be replaced with pre-fabricated filters, which will also drive down complexity significantly. The potential of using an ASE source for interrogation is particularly attractive, however more studies are required to investigate the impact of noise on the reconstruction. 

The complexity has now shifted from the device-end to the computational end. This shift in the measurement paradigm is also evidenced by more recent work in this area \cite{wang2019single,kita2018high}. This is advantageous, as it is possible to adapt and improve the optimization algorithms used to speed up acquisition, processing and accuracy of the CS-data, potentially without changing the optical hardware. As we are using a simple optimization modality (i.e. $l_1$-minimization), the computing power required is very low, and hence the reconstruction code can be ported onto widely available microcomputer platforms like Raspberry PIs, which have shown their viability for similar or even more computationally intensive tasks\cite{chagas2017100, aidukas2019low, 10.1117/12.2309763}.

At present, the demonstrated performance and associated accuracies are at best modest. For instance, while our values for centroid variance is $\sim$10-20 pm, conventional optical spectrum analyzers and commercial interrogators offer much higher precisions. Also, there are challenges in adapting this approach for interrogating multiplexed gratings - a task which is performed by the recently demonstrated incoherent OFDR-based methods \cite{WerzConf, WerzMDPI}. Also, OFDR based-methods intrinsically offer much higher spatial and spectral resolutions.  Yet, even in its current restrictive form, the CS-based interrogator offers an alternative to existing state-of-the-art interrogators, striking a trade-off between performance and complexity, and it can be of interest in applications (e.g. weather monitoring, environment sensing) where high-speed ($\sim$kHz) or high-resolution ($\sim$pm) is not paramount. Both the precision and the acquisition rate are currently limited by the programmable filter used, and also by limits imposed by the reconstruction algorithm and its associated inaccuracies. Precision limitations can be addressed by use of advanced fitting routines beyond centroid detection, and advanced optimization algorithms, while bottom-up engineering approach to the interrogator design can be used to optimize its dynamic performance. These hardware and software limits can be lifted by making the most of the current boom of machine learning and AI-based research and its associated enabling technologies, making the way for low-complexity devices, which can be mass produced, 
making them more accessible for wider adoption. 

\medskip

\medskip
\textbf{Acknowledgements}

SS would like to thank Dr. Vladimir Gordienko and Dr. Nand Kumar Meena for fruitful discussions. This project has received funding from the European Union’s Horizon 2020 research and innovation programme under the Marie Skłodowska-Curie grant agreement No 713694 (MSCA-COFUND-MULTIPLY). 

\medskip
\textbf{Disclosures}

The authors declare no conflicts of interest. 

\medskip
\textbf{Data availability statement}

The data that support the findings of this study are available from the corresponding author upon reasonable request.

\bibliographystyle{IEEEtran}
\bibliography{IEEEabrv,Bibliography}

\end{document}